# Lifting restrictions on coherence loss when characterizing non-transparent hypersonic phononic crystals

Konrad Rolle,[1] Dmytro Yaremkevich,[2] Alexey V. Scherbakov,[2,3] Manfred Bayer,[2,3] and George Fytas[1,*]

[1]Max-Planck-Institute for Polymer Research, Ackermannweg 10, 55128 Mainz, Germany

[2]TU Dortmund, Experimentelle Physik 2, 44227 Dortmund, Germany

[3]Ioffe Institute, 194021 St. Petersburg, Russia

*Corresponding author: fytas@mpip-mainz.mpg.de

**Abstract.** Hypersonic phononic bandgap structures confine acoustic vibrations whose wavelength is commensurate with that of light, and have been studied using either time- or frequency-domain optical spectroscopy. Pulsed pump-probe lasers are the preferred instruments for characterizing periodic multilayer stacks from common vacuum deposition techniques, but the detection mechanism requires the injected sound wave to maintain coherence during propagation. Beyond acoustic Bragg mirrors, frequency-domain studies using a tandem Fabry-Perot interferometer (TFPI) find dispersions of two- and three-dimensional phononic crystals (PnCs) even for highly disordered samples, but with the caveat that PnCs must be transparent. Here, we demonstrate a hybrid technique for overcoming the limitations that time- and frequency-domain approaches exhibit separately: We inject coherent phonons into a non-transparent PnC using a pulsed laser and acquire the acoustic transmission spectrum on a TFPI, where pumped appear alongside spontaneously excited (i.e. incoherent) phonons. Choosing a metallic Bragg mirror for illustration, we determine the bandgap and compare with conventional time-domain spectroscopy, finding resolution of the hybrid approach to match that of a state-of-the-art asynchronous optical sampling setup. Thus, the hybrid pump-probe technique retains key performance features of the established one and going forward will likely be preferred for disordered samples.

**Introduction**

Phononic crystals (PnCs)[1,2] can manipulate sound at various frequencies, yet the hypersonic (GHz) regime is particularly interesting because corresponding acoustic wavelengths are commensurate with that of light. PnC research can then build on existing work on photonic crystals (PhCs)[3,4], and appropriately designed dual PnC-PhC ("phoXonic") structures[5,6,7,8] can enable controlled photon-phonon interaction for signal processing and communication devices. However, another motivation for working at hypersonic frequencies is the availability of sophisticated optical spectroscopy techniques for characterization. Thus, the earliest studies of bandgaps for high-frequency phonons pertain to one-dimensional (1D) Bragg mirrors[9,10] and hypersound spectroscopy by pulsed pump-probe laser systems was applied[11,12] to these structures almost as soon as the corresponding "picosecond ultrasonic" (PU) technique[13] had been demonstrated. This method however requires that the sample contain at least one non-transparent (typically, metallic or semiconducting) transducer layer where the pump pulse excites phonons. Due to this constraint, the technique has been more widely adopted for characterizing inorganic Bragg lattices than for e.g. polymer samples[14,15].

Since PU samples the acoustic excitation in the time-domain, another drawback is that the sound wave propagating through the PnC not only suffers dissipation, but has to maintain coherence as well. Indeed, for very high frequencies (> 100 GHz) where vibrations are responsible for heat transport, surface roughness is a cause of diffusive scattering during phonon propagation that manifests itself as an increase in thermal boundary resistance[16,17]. The same effect is known to adversely impact PU experiments[18,19] but is typically not an issue when working with Bragg mirrors manufactured by common vacuum deposition techniques for solid state devices. However, soft-matter as well as higher dimensional hypersonic PnCs are also of topical interest for e.g. cavity optomechanics[20,21] and often naturally exhibit partial disorder as a result of the manufacturing process, which has been linked[22] to decoherence in the two-dimensional (2D) case. For three-dimensional (3D) structures, colloidal self-assembly of silica spheres was used to make opal (face-centered cubic, fcc) PnCs[23,24,25] but time-domain pump-probe experiments proved challenging, since the acoustic transmission spectrum could not be acquired directly. Rather, the cited experiment relied on detection of reflectivity variation at the attached metal transducer, in the vicinity of the first few opal layers and thus, prior to phonon decoherence through propagation. However, this strategy requires the probe pulse to penetrate the PnC, which would arguably be impossible for an opaque PnC and accordingly places strong hypotheses on PnC photonic properties. While opals from silica spheres form a transparent thin film for low layer numbers (< 20) not only does this not hold true for all PnCs, but eliminates the case for using pump-probe techniques altogether.

Indeed, since for transparent PnCs it is possible to probe inside the structure directly, one can roundly dispense with injection and detection of coherent sound waves. Instead of taking a time-domain approach, it is easier to work in the frequency-domain and use an interferometer, which can detect light inelastically ('Brillouin') scattered from thermally excited (i.e. incoherent) phonons. Although for GHz frequencies the corresponding instruments (tandem Fabry-Perot interferometers, TFPI)[26] became commercially available around the same time as the first PU experiments, their application to hypersonic phononic bandgap characterization[27,28,29] had to wait for progress in fabrication techniques to enable two- and three-dimensional hypersonic PnCs, such as the opals just mentioned. This is not only because there are no real advantages of so-called frequency-domain 'Brillouin light scattering' (BLS) over time-domain techniques for characterizing one-dimensional Bragg mirrors: For bulk phonons, scattered frequencies are selected according to the Bragg condition, so that detection has to be angle-resolved if frequency is to be scanned. Even then, the law of refraction limits the range of beam propagation angles accessible inside the sample. Thus, frequency scans are truly broadband only when in-plane order is probed. Access to out-of-plane order however is more limited[30] though a prism[31] can widen bandwidth (note that in-plane order is generally inaccessible in the time-domain, due to transducer film geometries). Nevertheless, the advantage of angle-resolved experiments (which in principle are also possible in the time-domain) is that full PnC dispersion relationships can be obtained. By contrast, PU studies of non-transparent PnCs (even for the simple 1D case) are limited to observing the bandgap in the acoustic transmission spectrum indirectly through reflectivity variation.

Though time- and frequency domain approaches have been successful separately on non-transparent Bragg lattices and transparent colloidal PnCs respectively, in between there is a wide unexplored field of e.g. colloidal crystals assembled from non-transparent particles[32], or inverse opals of air-spheres in metal[33,34]. To bridge this gap, we shall here introduce a hybrid pump-probe technique, where phonons are injected into a PnC by a pulsed pump laser as in PU, but detected by a TFPI as in BLS, i.e. by analyzing the frequency-shift of continuous-wave (CW) probe light scattered from phonons transmitted by the PnC. Previous proof-of-principle experiments have already combined pulsed[35,36] and even harmonic[37] phonon pumps with BLS, where the samples were rectangular prisms that had one facet coated with a metallic transducer film. Conceptually, it is then straightforward to place the PnC between transducer and detection prism, and scan different frequencies by changing scattering angle by means of a goniometer. Since colloidal self-assembly techniques often produce spatially heterogeneous specimens, there is however an interest in keeping laser spot-size small. This in turn is incompatible with typical focal lengths in a goniometer and requires another approach, as discussed in the next section. For the sample, while 'non-transparent opals' would be the most obvious in terms of applications, our focus here is on demonstrating and evaluating the technique. Thus, a metallic Bragg lattice shall be explored in this study as arguably the simplest non-trivial sample, permitting us to validate the hybrid technique by subsequent comparison with results from time-domain methods.

**Experiment**

Since the parameters of the experimental setup constrain the sample design more than the other way round, the former shall be discussed first. For phonon generation, ultrafast lasers offer nearly unlimited bandwidth, the depletion of the probe beam by Brillouin scattering being so low that there is generally little benefit to using longer pulses[38] or a harmonic pump[37]. For hypothetical use with a goniometer, a high energy (µJ) pulse source is advantageous[35], since energy concentration in time relaxes constraints on energy concentration in space for a similar signal-to-noise (SNR) target ratio (assuming appropriate time-gating of the TFPI detector[35]). Thus, larger spot-sizes compatible with practical focal lengths of goniometer optics can be used[39], which also permits for lower incidence angles, as well as reducing heating and acoustic beam diffraction. When high spatial resolution is desired however, a smaller spot can be achieved by using a high repetition rate femtosecond oscillator, where low pulse energies avoid ablation. The latter case is realized for the pump laser in the present study (Toptica femtofiber ultra, 780 nm, 500 mW, 80 MHz, 6 nJ/pulse). The required spot-size then depends on the transducer material; for broadband phonons that are needed to measure bandgaps, use of aluminum (Al) for generation[39] avoids acoustic impedance mismatch if a glass substrate is used on the detection side to observe the PnC transmission function (for exciting frequencies < 10 GHz, the bipolar pulse generated at the Al/air interface may not be suitable[40], so that e.g. a polymer film has to be spin-coated on top). For this transducer material, spot-size should be < 5 µm[36], rendering the traditional goniometer approach impractical.

For achieving low angles of incidence, an alternative to a prism is use of an oil-immersion objective, where total internal reflection fluorescence ('TIRF') has made numerical apertures (NAs) of up to 1.49 widely available. For frequency-domain Brillouin studies, oil-immersion objectives[41,42] are sometimes used in backscattering geometry (fixed scattering angle at 180°), after analysis dispelled initial worries about excessive linewidth broadening at high NAs[43]. They have also recently been used with off-axis beam injection for frequency-domain studies of in-plane anisotropy[44], albeit also with a 180° fixed scattering angle. Here, we propose to use a high NA objective to scan over a wide range scattering angles and hence, frequencies. The setup is shown in Fig. 1: The sample is a PnC coated onto a microscope glass coverslip. The pump laser is focused by an aspheric lens (Thorlabs, A375TM-C, f=7.5 mm) to a micrometer-sized spot on the Al-transducer film vacuum-deposited onto the PnC, where large film thicknesses (> 200 nm) are necessary to avoid heat buildup. Phonons are generated in the transducer and propagate through the PnC into the glass coverslip. On the probe side, rather than use conventional backscattering (thin green line through greyed-out elements) we inject the beam off-axis into our objective (Zeiss, Plan-Apochromat, 63x, NA=1.4, 440762-9906) by means of a knife-edge mirror (Edmund Optics, 36-137) on a translation stage. With the latter, beam distance from entrance pupil center and hence scattering angle can be roughly controlled. This is similar to how the impulsive stimulated Brillouin scattering ('transient grating') method[45,46] uses (for studies of dispersion-relationships, changeable[47]) phase-masks to inject a probe beam off-axis into the focusing optics. Inside the glass coverslip, both spontaneously excited and coherent phonons are present, but the former are isotropically distributed while the latter have a well-defined direction (normal to the transducer) and polarization (longitudinal). Thus, for the (much stronger) coherent phonons, the Bragg condition is only fulfilled for one set of angles (frequencies), obviating use of masks (as for spontaneous phonons[44]) for selecting the angle of the scattered light. Note however that the latter strategy would not be well-suited to characterizing transparent PnC dispersions since use of the objective (as compared with that of a goniometer) sacrifices precise knowledge of the incidence angle. Here, this information would only amount to the dispersion relationship of the glass substrate, so that only the transmission spectrum of the non-transparent PnC is finally accessible.

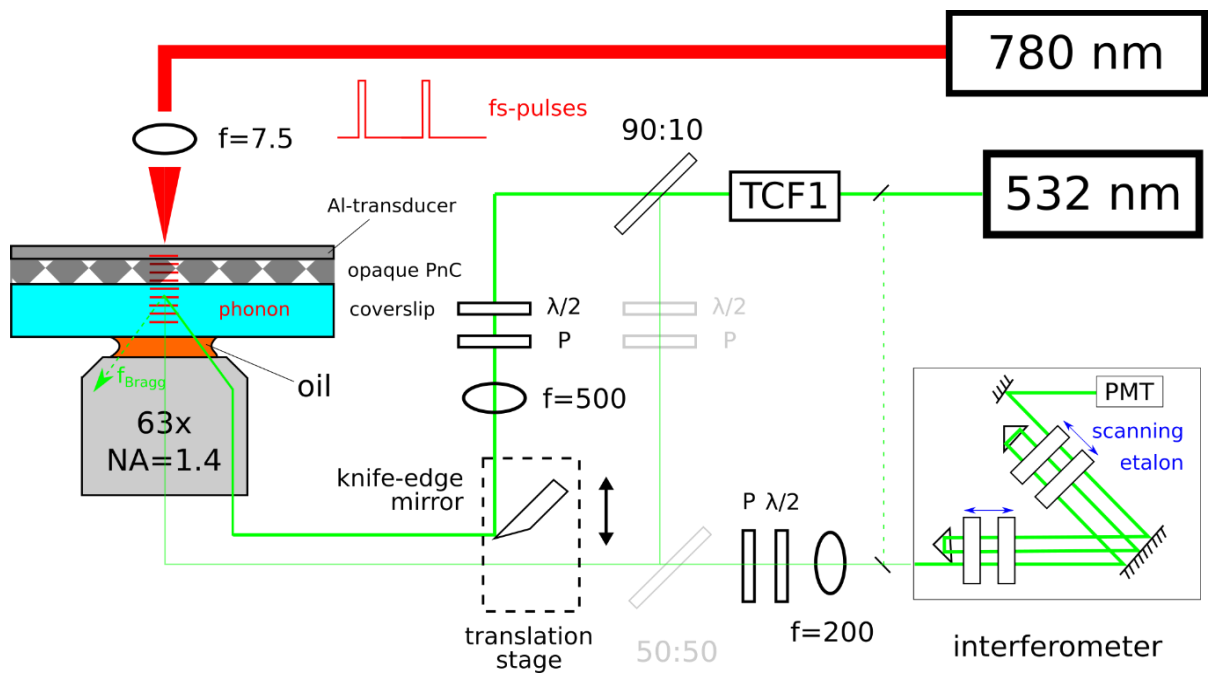

**Figure 1.** Setup for hybrid pump-probe experiments on a non-transparent phononic crystal (PnC), where the probe is a narrow-linewidth continuous-wave laser ("532 nm") and the pump a femtosecond oscillator ("780 nm"). Broadband longitudinal phonons are injected into the PnC (diamond pattern) by the pump beam (thick red line), which is focused by an aspheric lens (f=7.5 mm) to a micrometer-sized spot on an aluminum (Al) transducer film. Propagating into the coverslip, they form a moving diffraction grating by which the probe beam is scattered and frequency-shifted according to its angle of incidence on the grating. Normal incidence (thin green beam path) is always possible but lower scattering angles are achieved through use of a high numerical aperture oil-immersion objective ("63x, NA=1.4"). Here, frequency is controlled by offsetting the focused (f=500 mm) probe beam (thick green beam path) from the center of the objective entrance pupil by means of a knife-edge mirror on a translation stage. Polarizers ("P") and half-waveplates ("λ/2") configure the beampath for longitudinal phonon analysis, while a mode-cleanup filter ("TCF1") suppresses out-of-band laser noise. Finally, scattered probe light is analyzed by means of a tandem Fabry-Perot interferometer.

As mentioned, working with small spot-size imposes a low-frequency cutoff, an issue exacerbated by the finite NA of the objective. Thus, for NA=1.4, the corresponding 2θ-angle on a goniometer is only 45.8°. Knowing the wavelength of the probe laser (SpectraPhysics Excelsior, 532 nm, 300 mW) we can then calculate the lower cutoff frequency for glass coverslips (n=1.52, c=5660 m/s) to be $f_{min}=(2nc/\lambda)*sin(\vartheta)$=12.6 GHz while the high-frequency cutoff is always $f_{max}=(2nc/\lambda)$=32.3 GHz (with NA=1.49, $f_{min}$=6.6 GHz). Sample design has to be consistent with these limits: Accordingly, we choose a center frequency of $f_0$=22.5 GHz for the Bragg mirror, with titanium ($\rho_{Ti}$=4500 kg/m$^3$, $c_{Ti}$=6100 m/s, $Z_{Ti}$=27.49 MRayl) and silver ($\rho_{Ag}$=10490 kg/m$^3$, $c_{Ag}$=3650 m/s, $Z_{Ag}$=38.29 MRayl) as the constituents. Then, the dispersion can be calculated analytically for an infinite quarter-wave lattice[48,49], whence a bandgap width of $\Delta f=(4/\pi)*f_0*asin((Z_{Ti}-Z_{Ag})/(Z_{Ti}+Z_{Ag}))$=4.72 GHz which is well-matched to the observation window $f_{max}-f_{min}$ accessible by the objective. This can also be seen in Fig. 2 from a transfer matrix simulation of the acoustic transmission function for the case of a finite (10 layer) lattice.

With all of these considerations in mind, metallic Bragg mirrors with 10 periods of alternating titanium and silver layers are fabricated on microscope glass coverslips and diced silicon wafers (Ted Pella) as substrates. Layer thicknesses ($d_{Ti}$=68 nm and $d_{Ag}$=40 nm) are set to match the acoustic quarter-wavelengths corresponding to the target $f_0$. Glass substrates are first cleaned by ultrasonication in Hellmanex solution, then washed multiple times with Milli-Q water and dried overnight. Deposition is carried out at Fraunhofer IMM (Mainz, Germany) in a fully automated e-beam evaporator (Balzers Bak 640). Coverslips are fixed by small neodymium magnets (supermagnete.de, S-01-01-N) placed in their centers to the chamber steel ceiling, which is then pumped to a vacuum of 2*10$^{-6}$ mbar. After an initial Argon cleaning, the 20 alternating layers are deposited without breaking vacuum, starting with titanium for the first layer. The final silver layer is capped by a thick (250 nm) Al film. Despite the large overall structure thickness, the Al layer (for glass substrates) is found to be perfectly reflecting after opening the vacuum chamber. Finally, Bragg mirrors deposited on silicon substrate are imaged by scanning electron microscopy (SEM) after a trench has been cut on an ion-milling system (Fig. 2 inset).

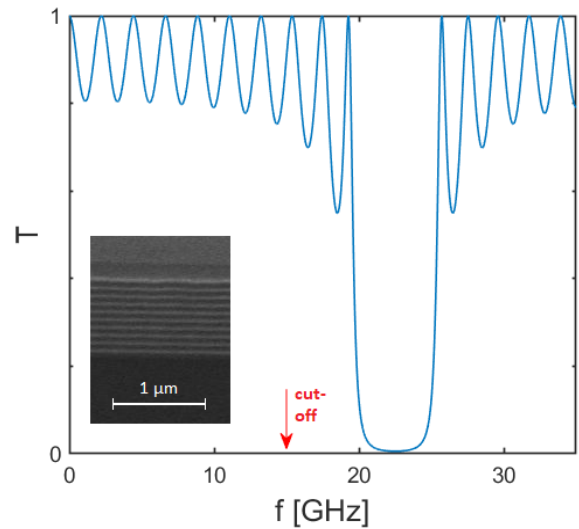

**Figure 2.** Acoustic transmission function of the Ti/Ag-Bragg lattice sample (10 periods) simulated for the target quarter-wavelength dimensions by the transfer matrix method, where the red arrow indicates the approximate lower cut-off frequency of the NA=1.4 objective used in the present study (inset: scanning electron microscopy image of the manufactured sample)

For the hybrid pump-probe experiment, the coated glass coverslips are mounted perpendicular to the objective in inverted microscope configuration. Since the metallic lattice acts as a mirror, the normal orientation leads to a lot of elastically (Rayleigh) scattered light being present in the return beam path (at the same time, the small pump spot-size entails high acoustic beam divergence, thus limiting achievable defocusing from the excitation region). Though the interferometer (JRS TFP2-HC) is equipped with a shutter unit (JRS LM-2) at its entrance that rejects elastically scattered light, the latter can also contain significant out-of-band laser noise. Hence, a filter (JRS TCF-1) is installed in the path of the probe laser. Starting out in standard backscattering configuration (thin green beam path, Fig. 1), the coverslip is lowered onto the objective until the focus is on the metal coating and then translated horizontally until the transparent window left in the Bragg mirror after removal of the neodymium magnet is found. By observing both pump and probe foci on a camera (JRS PC-1), the pump is brought to coincidence with the probe by translating the asphere. Then, to verify presence of pumped phonons in backscattering, the spectrum is displayed in real-time on the TFPI software for both "pump on" and "pump off" states, whose free spectral range (FSR) has been set to 34.5 GHz (> $f_{max}$). Finally, the 50:50 beamsplitter for backscattering (greyed out in Fig. 1) is removed and the knife-edge mirror translated 2 mm in steps of 0.1-0.25 mm to acquire different parts of the PnC transmission spectrum while the pump is on. For reference, a blank sample of aluminum (> 200 nm) deposited by a benchtop sputter coater (Baltec MED-020) onto a glass coverslip was also characterized in the same manner.

Since the main interest of the Bragg lattice sample is ease of comparison with the established PU technique, we finally characterize with a time-resolved pump-probe approach. No additional thought has to be given to sample design because these experiments can achieve very large bandwidth by probing reflectivity variation at interfaces rather than bulk phonons. Since delay-lines cannot be made infinitely long, traditional time-domain pump-probe setups that incorporate them are however unlikely to be competitive in terms of frequency-resolution with a TFPI. Thus, an asynchronous optical sampling (ASOPS) setup[50] is chosen for characterization, where frequency-resolution[51] is only limited by pulse repetition rate and acoustic losses in the sample. Two femtosecond oscillators similar to those in the hybrid pump-probe experiment we just described are used for the probe (Toptica femtofiber ultra 780) and pump (Toptica femtofiber ultra 1050), which are locked to a repetition rate difference of 1600 Hz (> 1 THz bandwidth) by a synchronization unit (Laser Quantum TL-1000-ASOPS). Because again oscillator-type lasers are employed, all the aforementioned considerations on spot size apply as well; in a PU context, the resulting high spatial resolution has been shown to be sufficient for bioimaging applications[52]. Probe power was kept constant at 15 mW while three different measurement configurations were trialled. In reflectance pump and probe were focused through the same microscope objective (Mitutoyo M Plan Apo NIR 20X) onto the Al-coated side of the sample, while in transmittance the pump beam was focused by the 20x objective, and the probe by a corresponding 50x objective. As for the hybrid pump experiment, the transparent window left after removal of the neodymium magnet in the sample center was used as an alignment help for bringing pump and probe foci to coincidence in the latter configurations. Though both possible sample orientations were explored, best results corresponded to time-domain traces acquired for the pump focused at rather low power (7.5 mW) to a spot of 2 µm diameter (FWHM) onto the titanium layer (excitation density of 3 mJ/cm$^2$) and the probe to a spot of 1 µm diameter on the side of the Al/air interface. Similarly to the hybrid case, a broadband (> 100 GHz) wavepacket of coherent longitudinal phonons is then generated by pump-induced ultrafast heating and thermal expansion, but detected after propagation through the phononic Bragg mirror by the probe which monitors the modulation in Al reflectivity due to the elasto-optical effect[13]. Finally, the fast Fourier transform (FFT) of the acquired time-domain data was computed after subtraction of the exponentially decaying thermal background and filtering were performed.

**Results and discussion**

In order to have a valid point of comparison, it is instructive to first examine spectra from application of the hybrid method to a blank sample (Fig. 3a). For the latter, phonon generation and detection also take place at an Al-transducer and a glass coverslip respectively (upper left inset of Fig. 3a), but no PnC is placed between both to filter out selected frequencies. In conventional backscattering configuration, PnC presence is not expected to have an effect, since spectra from pump-enhanced BLS (light grey) show narrowband peaks that differ in intensity but not qualitatively from those produced by spontaneous Brillouin scattering alone (not shown) in the "pump off" state. Indeed, only some slight asymmetry between Stokes and anti-Stokes (respectively, frequency down- and upshifted) scattered light is observed. The situation is radically different when the probe beam is not injected in the objective center, but at various off-axis positions by translating the knife-edge mirror. For each position (circular inset of Fig. 3a) a corresponding spectrum is acquired. For spontaneous Brillouin scattering, a larger range of angles and hence frequencies is probed, leading to a considerably broadened plateau-like response, barely visible around -30 GHz and identical in all three spectra. It is dwarfed however by the signal from the pumped phonons, manifested as a peak that moves to higher frequencies as the knife-edge mirror is brought closer to the center. As part of this trend, it also becomes sharper and more intense, possibly indicating increased light collection efficiency of the objective near the entrance pupil center. The overall spectral broadening is a result of the choice of focusing optics (f=500 mm) that involves a trade-off against signal-to-noise ratio. It indicates a wide angular distribution for incident light, but should be an asset when less sharply resolved PnC spectral features have to be examined. Also, the range over which frequency can be scanned is found to correspond reasonably well to the above prediction, with $f_{min}$ ~ 15 GHz for NA=1.4 not far from the 12.6 GHz calculated above. Compared with central injection of the beam, the most remarkable difference is however not peak broadening but asymmetry between Stokes and anti-Stokes intensities. While Fig. 3a shows probe light to be only upshifted in frequency when scattering from coherent phonons, in practice, it is quite sensitive to alignment. Thus, the acoustic beam can cross incident probe light 'before' or 'after' it gets reflected at the Al-mirror constituted by the transducer bottom. In the latter case, subtraction of the vectors defined by the pump and light propagation directions downshifts scattered light frequency (since it also changes its direction, a 2$^{nd}$ reflection at the Al-mirror occurs). Thus, even though pumped phonons (unlike spontaneously excited ones) have only one propagation direction, in practice, pump-enhanced spectra show varying asymmetry, with backscattering spectra being close to symmetric (light grey in Fig. 3a).

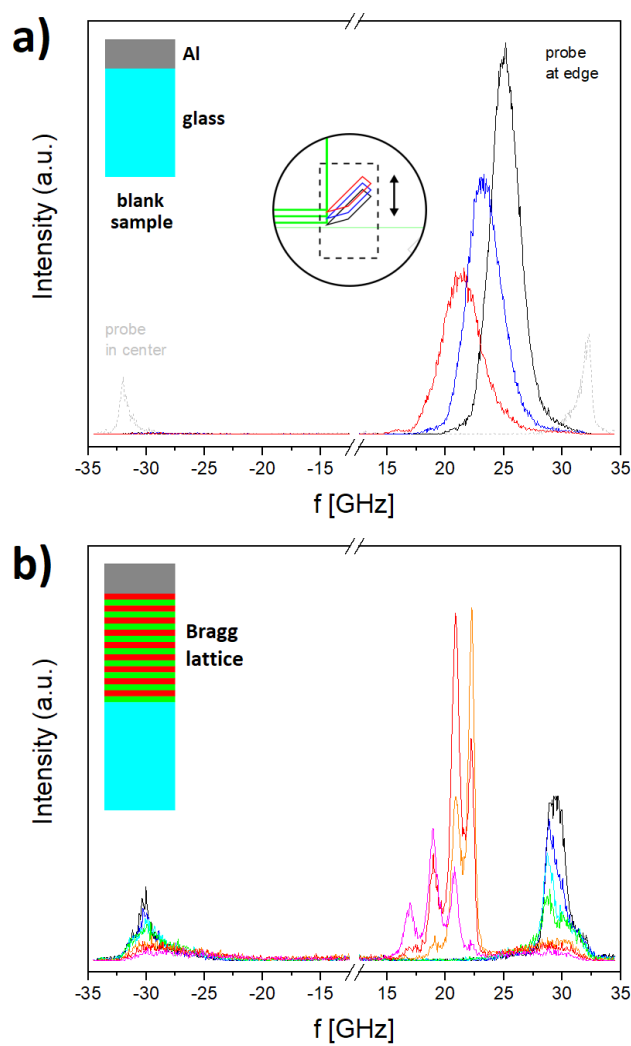

**Figure 3.** Hybrid pump-probe spectra, with inset in the upper left corner illustrating sample structure a) blank sample (glass with > 200 nm Al-film), showing the effect of knife-edge mirror translation (circular inset) on frequency and intensity of the scattered light, with higher frequencies corresponding to positions closer to the entrance pupil center (dashed grey spectrum shows pump-enhanced backscattering i.e. thin green beam path in Fig. 1); differential screw interval 0.25 mm and acquisition time 30 s for spectra from the edge-injected beam (backscattering spectrum not to scale) b) transmission function of Bragg mirror (as in Fig. 2) for various knife-edge mirror positions; differential screw interval 0.25 mm for spectra showing transmission below the bandgap, and 0.125 mm for spectra showing transmission above the bandgap; acquisition time 5 min for all spectra

Characterization of the Bragg lattice is conducted in the exact same manner as for the blank sample, with data presented in Fig. 3b according to the same format. As before, several spectra corresponding to different knife-edge mirror positions are plotted on top of each other. Again, there is a (symmetric) background signal from spontaneous phonons around 30 GHz, whose slightly sharper response for more central knife-edge mirror positions may be due to the more restricted angular range. More importantly however, for the pumped phonons, a non-trivial response is observed now, with the envelope of all spectra taken together effectively giving the PnC transmission function. At lower frequencies, peaks with interference minima are observed, which stay constant between individual spectra even though relative intensity of peaks at higher frequencies increases as higher incidence angles are scanned. The signal is seen to disappear around 22.6 GHz at what obviously constitutes the lower bandgap edge, and reappears gradually above it. Above the bandgap edge, no interference minima could be found even scanning beyond the shown range, which might be due to the fact that pump response at higher frequencies becomes narrower (see also Fig. 3a). Also, deviations of layer thickness respective to quarter-wavelength condition can result in transmission dips that are less deep on one side than on the other, confirmed by the upper bandgap edge being less steep than the lower. Mainly however, it lies perilously close to the cut-off given by $f_{max}$, rendering observation of interference fringes difficult. Indeed, the whole spectrum is shifted upwards in frequency respective to what could have been expected from the simulation in Fig. 2 on which the original design was based. Thus, the mid-frequency is found at $f_{0,meas}$=25.6 GHz rather than $f_{0,design}$=22.5 GHz, indicating a lattice constant that is smaller than intended, though the width of ~ 5 GHz is still very close. As a consequence of this upshift, three interference minima at lower frequency fall into the observation window. Spacing between the nodes is uniform at ~ 2 GHz both in the simulation and measurement, leaving no doubt about the number of periods being correct. Finally, the result allows for little ambiguity in its interpretation.

A hybrid pump-probe approach is not expected to offer decisive advantages over established PU spectroscopy for one-dimensional PnCs. Nevertheless and in view of the very clear spectrum also, this kind of sample lends itself very well to a performance comparison between both techniques, which might also help to dispel lingering doubts about the discrepancy from simulation. Hence, data from PU is presented in Fig. 4, where a 'transmittance' spectrum was selected among the three above-described configurations (illustrated for the chosen spectrum in the inset on the left panel) because of its closest resemblance to the hybrid pump-probe experiment. Nevertheless, data from the other two measurements (not shown) yielded similar information, a fact which underscores the versatility of PU whose use also relaxes any strong hypothesis on the substrate (since no oil-immersion objective is involved). Moreover, PU boasts huge bandwidth because it can not only measure Brillouin oscillations in the bulk but reflectivity variations at interfaces, though the simultaneous sensitivity to several effects can also complicate interpretation of the spectra. Thus, the transmission function obtained by computing the fast Fourier transform (FFT, right panel) of the transient reflectivity signal (ΔR, left panel) not only shows a pronounced dip corresponding to the primary bandgap around 25 GHz, but also one at 76 GHz corresponding to its higher-order replica. Moreover, due to increased bandwidth destructive interference dips above the bandgap edge from the finite period number of the Bragg mirror can now be confirmed (again, they appear somewhat less deep and the edge less steep than for the lower frequencies). Both above and below the bandgap, interference dips show lower contrast closer to the bandgap, which is contrary to the simulation but again seems to be consistent with the hybrid pump-probe results. Their spacing also agrees with the latter, and the nine oscillations in the frequency range below the bandgap can be confirmed from the calculated spectrum in Fig. 2. The presence of the broad peaks centered at about 31 GHz, 47 GHz and 85 GHz can be attributed to the initial spectrum of the strain pulse generated in the titanium by the laser pulse, affected also by the superlattice. Most importantly however, bandgap width and mid-frequency from hybrid and time-domain measurements agree almost perfectly, confirming that any discrepancy from simulation must be due to deviations of manufacturing or material parameters. Also in terms of frequency resolution, the linear region of the lower bandgap edge has a width of 300 [MHz] as measured by both techniques; in this regard and from the overall quality of the data, hybrid pump-probe spectroscopy is finally seen to be very much the equal of PU with an ASOPS setup. Thus, there does not seem to be any obstacle preventing application of the technique to more challenging samples from e.g. colloidal self-assembly, which have hitherto eluded the grasp of PU due to presence of disorder.

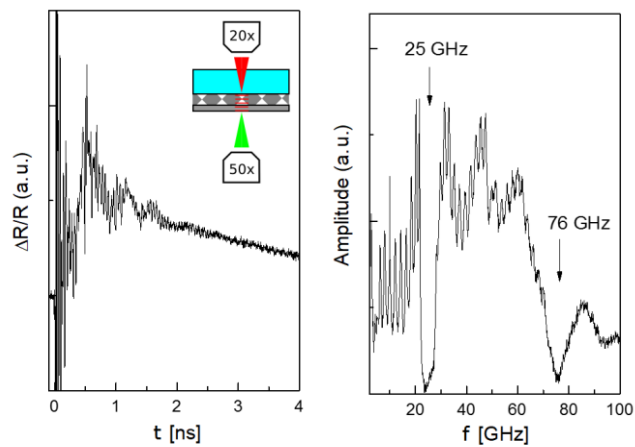

**Figure 4.** Picosecond ultrasonic data for the Bragg lattice sample after 10 minutes acquisition, in the time-domain (left) and after Fourier transform (right), where the upper-right inset schema from the time-domain frame illustrates the employed transmittance configuration (pump in red and probe in green, with the former incident on the final titanium layer and the latter on the Al-film)

**Conclusion and outlook**

We have presented a hybrid pump-probe technique for measuring hypersonic PnCs that lie outside the respective comfort zones of time-domain (picosecond ultrasonics, PU) and frequency-domain (Brillouin light scattering, BLS) GHz phonon spectroscopy. This method is based on combining a pulsed laser for phonon generation (as in PU) with a tandem Fabry-Perot interferometer (as in BLS). We have validated the approach by taking a sample (a metallic Bragg mirror) typically suited to a PU experiment and characterized it by both hybrid and time-domain pump-probe spectroscopy. Excellent agreement was found, with the new hybrid scheme already matching state-of-the-art time-domain pump-probe setups for both spatial and frequency resolution. However, due to restricted bandwidth notably, hybrid pump-probe spectroscopy is less versatile than the established approach. Hence, beyond possibly selected comparative studies with PU aimed at distinguishing decoherence and dissipation in the same samples, its use is likely to be reserved for structures that cannot be examined otherwise. Thus, studying e.g. 'non-transparent opals' would be the ultimate aim, since they lack both transparency (necessary for a BLS experiment) and are not expected to support phonon coherence (necessary for PU). However, polystyrene and silica spheres being most frequently used for colloidal PnC self-assembly, a transparent PnC might be more readily procurable for proof-of-principle, and offer the added advantage of allowing BLS characterization for comparison (note also that in an air-matrix, for > 20 layers, such opals become non-transparent due to photonic scattering and have thus only been examined after liquid infiltration[28]). Then, it is necessary to address issues with acoustic impedance mismatch (and heating by the pump laser) if e.g. polystyrene-based opals or other polymer systems are to be studied; typically, acoustic impedance matching layers such as graphite thin films would have to be explored. Also, we mentioned that the bandwidth in the hybrid pump-probe approach is restricted, but using different probe wavelengths (subject to interferometer restrictions) or detection substrates (such as lead glass coverslips) could render other frequency regions accessible. For the latter in particular, if the approach from the present paper with a small spot-size femtosecond oscillator is to be retained, it might be necessary to develop oil-immersion objectives that can work with higher refractive index glasses. This effort might however be justified since such materials are used for acousto-optic modulators, and achieving grazing incidence angles in them might therefore also benefit acousto-optic information processing applications by shifting down phonon frequencies to bandwidths where direct acquisition of the time-domain response on a photodiode and oscilloscope[53,54] is possible. Finally though, for spatially homogeneous samples, accessing frequencies < 10 GHz might call for a more traditional (though not necessarily simpler) approach[35] based on a pump laser with μJ pulse energies, combined with a prism mounted in a goniometer. Considering also increasing dissipative losses of polymers at higher hypersound frequencies, three-dimensional PnCs from non-transparent inorganics are thus likely to constitute the immediate application of hybrid pump-probe spectroscopy.

**Acknowledgements**

We thank Stefan Schmitt and Doris Ehrler at Fraunhofer IMM for manufacturing the Bragg mirror. We thank Anke Kaltbeitzel for lending us the Zeiss objective. We thank Maren Müller for SEM. We thank Gabriele Herrmann and Uwe Rietzler for helpful discussions about vacuum deposition. We thank Vitalyi Gusev and Pascal Ruello for helpful discussions about picosecond ultrasonics. We thank Rüdiger Berger and Andrey Akimov for helping identify collaborators for this work.

K.R. and G.F. acknowledge financial support by ERC AdG SmartPhon (Grant No. 694977). D.Y., A.V.S. and M.B. acknowledge financial support of DFG through CRR TRR142 (project A06).

**Author contributions**

G.F., K.R. designed and K.R. carried out the hybrid pump-probe experiment. M.B., A.V.S. designed and D.Y. carried out the picosecond ultrasonics experiment. K.R., G.F. and A.V.S. wrote the manuscript.

**Additional Information**

The authors declare no competing interests.